\documentclass[a4paper]{aa}  
\usepackage{txfonts}
\usepackage{graphicx}
\usepackage{natbib}
\bibpunct{(}{)}{;}{a}{}{,}

\newcommand{\bz}{\ensuremath{\langle B_z \rangle}}

\newcommand{\bs}{\ensuremath{\langle \vert B \vert \rangle}}

\newcommand{\te}{\ensuremath{T_{\rm eff}}}
\newcommand{\wda}{WD\,0011-721}

\newcommand{\wdc}{WD\,2138-332}

\newcommand{\ha}{H$\alpha$}

\begin{document} 
   \title{A new weak-field magnetic DA white dwarf\\ in the 
          local 20\,pc volume.\\
          The frequency of magnetic fields in DA stars}

   \author{ J.D. Landstreet\inst{1,2}
          \and
          S. Bagnulo \inst{1}
          }
   \institute{Armagh Observatory and Planetarium, College Hill,
   Armagh BT61 9DG, UK \\
              \email{john.landstreet@armagh.ac.uk,stefano.bagnulo@armagh.ac.uk;}
         \and
             University of Western Ontario, London, Ontario, N6A 3K7, Canada. 
             \email{jlandstr@uwo.ca}
             }
   \date{Received June 3, 2019; accepted Jun 28, 2019}

  \abstract{We report the discovery of a new magnetic DA white dwarf (WD), WD\,0011$-721$, which is located within the very important 20\,pc volume-limited sample of the closest WDs to the Sun. This star has a mean field modulus \bs\ of 343\,kG, and from the polarisation signal we deduce a line-of-sight field component of 75\,kG. The magnetic field is sufficiently weak to have escaped detection in classification spectra.  We then present a preliminary exploration of the data concerning the frequency of such fields among WDs with hydrogen-rich atmospheres (DA stars). We find that $20 \pm 5$\% of the DA WDs in this volume have magnetic fields, mostly weaker than 1\,MG. Unlike the slow field decay found among the magnetic Bp stars of the upper main sequence, the WDs in this sample show no evidence of magnetic field or flux changes over several Gyr. }

   \titlerunning{Discovery of nearby DA magnetic white dwarf}
   \maketitle
%

\section{Introduction}

Magnetic fields are present in about 10\,\% of white dwarfs (WDs). The observed field strength ranges from a few kiloGauss to almost a gigaGauss in different WDs, a span of more than five orders of magnitude. Generally, the observed fields have a globally organised structure, with large areas of emerging flux on one hemisphere, and entering flux on the other. Observations of variations of the mean line-of-sight magnetic field strength \bz, and sometimes of photometric stellar magnitude, reveal rotation periods that are typically between minutes and weeks \citep[e.g. ][]{Brinetal13,Garyetal13,Ferretal15,Landetal17}. 

The magnetic fields of WDs appear to currently be of fossil nature. They do not seem to be maintained by any current dynamo action, but are probably the remnants of fields generated during earlier stages of stellar evolution. They have survived very slow Ohmic decay by virtue of the size and high electrical conductivity of the host star. The origin of these fields is still very uncertain. They may be generated as a result of dynamo action in the convective core during an earlier stage of stellar evolution, or they may be the result of the intense interaction during a binary system merger \citep{Toutetal08,Ferretal15}. There may be multiple evolution paths leading to presence or absence of a global magnetic field in the final WD state. 

It is worth noting that, on the whole, WDs are quite uninformative about their previous evolution. The majority show the spectral lines of only one element, hydrogen, in their optical spectra. A significant number show no spectral lines at all. From spectroscopy, it is often possible to derive accurate values of mass, radius, effective temperature, luminosity, and cooling age. However, very little further information can be extracted from spectra. In principle, magnetic fields can break this degeneracy because only a fraction of WDs carry detectable fields, and these fields vary widely in strength and probably in structure. Once we understand how to read the information contained in them, the fields should thus provide really valuable clues about physical processes that occurred in previous evolution.  It is certainly worthwhile to try to understand how these fields develop, evolve, and reflect earlier evolution. 

A solid observational description of which individual WDs have fields is fundamental to understanding the physical processes that these fields may reveal. A comprehensive description includes what fraction of WDs host these fields, how the observed fields are related to WD initial and present mass, atmospheric chemistry, and age since contraction to the WD state.  Although WD magnetism has been known for almost 50 years, information about all of these characteristics is still quite fragmentary. 

In order to characterise the global qualities of magnetic white dwarfs (MWDs), and to identify their relationships to other WDs and to the larger framework of stellar evolution, it is useful to study a volume-limited sample of WDs. Since the great majority of stars evolve to a WD final state, such a sample represents a kind of time capsule, which encodes in a direct way the results of more than $10^{10}$\,yr of star formation and stellar evolution in our region of the Milky Way galaxy. 

The largest volume-limited sample of WDs that has been extensively studied is the sample currently lying within 20\,pc of the Sun \citep{Holbetal16,Holletal18}. This volume of space is large enough to contain well over 100 WDs, enough to provides useful statistical information about sub-samples. It is believed that WD membership in this sample is now very close to complete. Accordingly, we have been examining in detail the subset of this sample that are MWDs in order to identify systematic features of this sample. With such data, we can try to answer such questions as whether the presence or absence of a magnetic field, or MWD field strength or magnetic field structure, are related to WD atmospheric composition, to initial and final mass, or to cooling age. Such information should, in turn, help us to understand more clearly the evolution processes encoded in the observed magnetic fields. 

Magnetic data about the WDs in the 20\,pc volume sample are still extremely incomplete. Some of the WDs in the sample have been closely examined for magnetic fields over the years. A wide variety of fields have been discovered. However, for many of the WDs, the available data a few years ago were only sufficient to identify fields above about one MG, or they did not constrain the possible presence or absence of a magnetic field at all. In response, we have been actively observing WDs in this sample by a variety of methods in order to obtain a more complete sample of the MWDs present in the 20\,pc WD sample, and to obtain the strongest practical upper limits on the remainder. 

In the course of our survey we have already identified two new DA MWDs, WD\,2047+372, and WD\,2150+591, that are members of the 20\,pc sample \citep{Landetal16,LandBagn19}, and we have presented evidence that a DZ WD member, \wdc, may be magnetic \citep{BagnLand18}. In this paper we report the discovery of another DA MWD, \wda, which is also resident in this volume, and we describe its characteristics. We then carry out a preliminary assessment of the frequency of occurrence of magnetic fields among the DA WDs, based on examination of the sample of DA MWDs in the 20\,pc volume.

\section{Observations, reduction, and measurements}

The observations presented in this paper were obtained with the FORS2 instrument
\citep{Appetal98} of the ESO VLT. \wda\ was observed
with grism 1200R, covering the spectral range 5600--7300\,\AA,
with a 1\arcsec\ slit width, for a spectral resolving power of 2140 (spectral resolution $\Delta \lambda \approx 3.1$\,\AA). 
Our target was observed in spectropolarimetric mode
to measure circular polarisation (Stokes $V/I$) as well as the unpolarised 
spectrum (intensity, Stoke $I$).  Observations were
carried out using the beam-swapping technique \citep[e.g.][]{Bagetal09}
to minimise instrumental effects, and reduced as explained by
\citet{BagnLand18}. Magnetic field values were measured as explained in
the section below.
The log of the observations is given in Table~\ref{Tab_Log}.

\begin{table}
\caption{Parameters of the newly discovered DA magnetic WD}
\label{Tab_obs-stars}
\centering
\begin{tabular}{l r  }
\hline\hline
\multicolumn{2}{c}{ WD\,0011-721}    \\
\hline    			                                   				                  
Alternate name    & LP\,50-73       \\
$\alpha$ (J2000)  & 00 13 49.91     \\
$\delta$ (J2000)  & --71 49 05.03   \\
$\pi$ (mas)       & 53.23           \\
Johnson $V$       & 15.17           \\
Gaia $G$          & 15.05           \\
Spectrum          &  DA\,7.8        \\
$T_{\rm eff}$ (K) &  6340           \\
$\log g$ (cgs)    &  7.89           \\
age (Gyr)         &  1.66           \\
mass ($M_\odot$)  &  0.53           \\
\hline
\end{tabular}
\tablefoot{Data from \citet{Subaetal17} and Gaia Collaboration (\citeyear{Gaiaetal18})}
\end{table}
 
\begin{table*}
\caption{Magnetic measurements of newly discovered magnetic WD}
\label{Tab_Log}
\centering
\begin{tabular}{l l l l c c r r@{$\pm$}l r@{$\pm $}l}
\hline\hline
Star          & Instrument & Grism      &  MJD        &\multicolumn{2}{c}{Date} & Exp.&
\multicolumn{2}{c}{\bz}&\multicolumn{2}{c}{\bs}\\
&            &            &             &  yyyy-mm-dd& hh:mm&
\multicolumn{1}{c}{(s)}&\multicolumn{2}{c}{(kG)}&\multicolumn{2}{c}{(kG)}\\
\hline
WD\,0011-721  & FORS2      &  1200R     &  58439.696  & 2018-11-17 & 04:42  &  2700  &$  74.9$&1.6& 343 &  15  \\[2mm]
              \hline
\end{tabular}
\end{table*}

\section{\wda}

\begin{figure}
\includegraphics*[width=9.3cm,trim=0.75cm 15cm 0cm  2cm,clip]{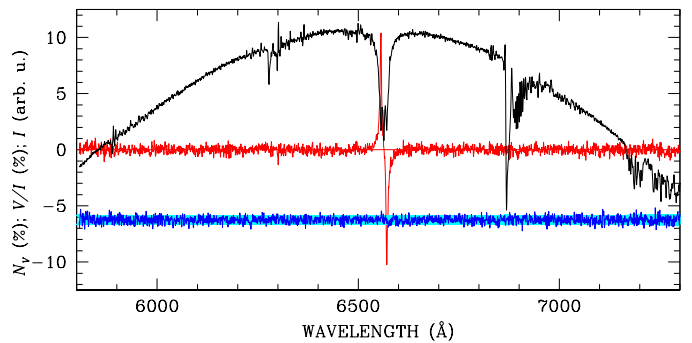}
\caption{\label{Fig_wda} 
  Polarised spectrum of \wda\ obtained with the FORS2 instrument with grism 1200R. The flux spectrum $I$ is shown in arbitarary units and not corrected for the instrument+telescope
  transmission function; the $V/I$ (red) and $N/I$ (blue) spectra are in percentage units, with $N/I$ shifted by -6\% for clarity. }
\end{figure}

\begin{figure}[ht]
\includegraphics*[width=9.3cm,trim=1.75cm 6cm 0cm 2.5cm,clip]{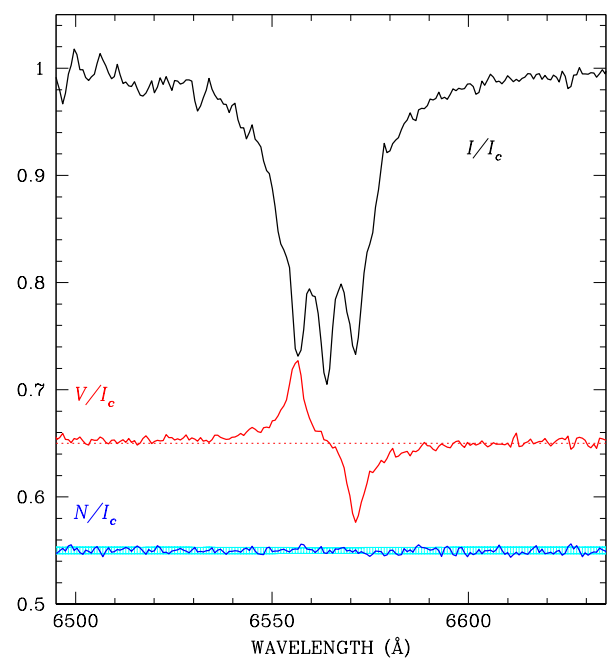}
\caption{\label{Fig_wda_detail} 
 Detail of polarised spectrum with continuum normalisation of \wda, obtained with the FORS2 instrument with grism 1200R. Here we show $I$ and $V$ (normalised to 1.0 in the $I$ continuum, with $V$ shifted upwards by +0.65 for clarity) rather than $I$ and $V/I$, in order to make clear the relative amplitude of the Stokes components $I$ and $V$.  The null profile is also shown offset by $+0.55$\,\%.}
\end{figure}

\wda is a cool DA with $\te \approx 6340$ \citep{Subaetal17} and is known to be a single star \citep[][and references therein]{Tooetal17}. In the optical spectrum, rather sharp lines of the H Balmer series are visible; no metal lines are detected. According to \citet{Subaetal17}, the star has a mass of $M = 0.53\,M_\odot$, slightly below the average of $0.6 M_\odot$ characterising most WDs. The main characteristics of this star are summarised in Table~\ref{Tab_obs-stars}.  A low-resolution optical spectrum from \citet{Giametal12} is available on the Montreal White Dwarf Database\footnote{http://www.montrealwhitedwarfdatabase.org} \citep[MWDD; ][]{Dufoetal17}. To our knowledge, there are no previous high-resolution observations available, nor any kind of polarimetric measurement; available low-resolution spectroscopy provides an upper limit to the field of the order of 1--2\,MG. 

Our single FORS2 polarised spectrum, obtained around \ha\ using grism 1200R, shown in Fig.~\ref{Fig_wda}, reveals both a clear Zeeman split line triplet in the core of \ha, and very strong circular polarisation of the $\sigma$ components of the line, significant at about the $40 \sigma$ level, indicating a substantial line-of-sight component of the field. The region immediately around \ha, rectified and normalised with a quadratic fit to the nearby continuum of Fig.~\ref{Fig_wda}, and expressed in units of $I$ and $V$ (rather than $I$ and $V/I$), in order to show clearly the relative amplitude of the two Stokes parameters) is shown in Fig~\ref{Fig_wda_detail}.

The presence of clear Zeeman splitting in the core of \ha\ allows us to measure the mean of the  magnitude of the surface magnetic field  {\bf B} averaged over the visible hemisphere, known as the mean field modulus \bs. When the field structure is such that there is a clear Zeeman triplet displaying three components of comparable strength and shape, a good estimator of this quantity is to measure the separation between the central, undisplaced $\pi$ component of the Zeeman triplet and the position of one the shifted $\sigma$ components. As discussed for example by \citet{Landetal15} and \citet[see Eq. ~(2)]{Landetal17}, \bs\ is related to the measured $\pi - \sigma$ separation $\Delta \lambda_{\rm Z}$ by 
\begin{equation}
    \Delta \lambda_{\rm Z} = 4.67\,10^{-13} g \lambda_0^2 \bs,
\end{equation}
where $\lambda_0$ is the undisplaced wavelength of the spectral line, $g$ is the Land{\'e} factor, equal to 1.0 for Balmer lines, and wavelengths are measured in \AA. Although the three Zeeman components are slightly asymmetric, the positions of the components of the line core, measured below the level where the three components merge, can be measured by eye or by fitting a Gaussian model to each component separately. The two methods agree well, and the mean $\pi - \sigma$ separation can be measured fairly accurately as $\Delta \lambda_{\rm Z} \approx 6.9 \pm 0.3$\,\AA, leading to $\bs \approx 343 \pm 15$\,kG.  This field is small enough that no significant correction is required for higher-order terms in the line splitting physics. Specifically, the quadratic Zeeman effect \citep[e.g.][]{Pres70} is negligible. 

Notice that for this star the value of \bs\ is quite well defined. The two $\sigma$ components of the Zeeman triplet have nearly the same FWHM as the central $\pi$ component, and in fact all three components have FWHM of slightly more than 3\,\AA. This width is considerably wider than the normal FWHM of the non-magnetic non-LTE core of \ha\ (about 1\,\AA), which is also the expected intrinsic width of the magnetically split central $\pi$ component. The observed widths of these three components  appear to be defined largely by the 3.1\,\AA\ resolution of the spectra. The compactness of the $\sigma$ components indicates that the local value of {\bf B} does not vary greatly over most of the visible hemisphere of the star. This compactness is confirmed by the Stokes $V$ spectrum; the regions of clearly non-zero $V$ are only a little wider than the apparent width of the $\sigma$ components in $I$. We estimate from the width of the $\sigma$ components that the intrinsic broadening is probably no more than about 2\,\AA, corresponding to a range of {\bf B} of only about $\pm 50$\,G over most of the visible hemisphere. 

The mean line-of-sight component of the magnetic field, averaged over the visible hemisphere at the time of observation, \bz, is obtained by computing the mean position of the spectral line as seen in right and left circular polarisation. The expression used for this measurement is 
\begin{equation}
    \bz \approx -2.14\,10^{12} \frac{\int v V(v) {\rm d}v}
                                    {\lambda_0 c \int [I_{\rm cont} - I(v)]{\rm d}v},
\end{equation}
where $v$ is wavelength measured in velocity units (cm\,s$^{-1}$) from $\lambda_0$, $I(v)$ and $V(v)$ are the measured intensity and circular polarisation Stokes components as functions of $v$, and $I_{\rm cont}$ is the continuum level relative to which the shifted right and left polarised line component positions are measured \citep{Math89,Donaetal97}. In most (hotter) weak-field magnetic DA WDs, Zeeman polarisation is detected far more strongly in the deep NLTE line core than in the very broad line wings, and the integral of Eq.\,(2) is measured using only the line core. In the case of \wda, however, the \ha\ line resembles a strong metal line, almost without important extended wings, and the entire line profile can be used to measure accurately the shift of the line centroid between its position as seen in left and in right circular polarisation. In this case we take the value of the rectified continuum, $I_{\rm cont}$, to be 1.0 (as in Fig.~\ref{Fig_wda_detail}). 

Using Eq.\,(2) and following the discussion of \citet{Landetal15} to estimate its uncertainty, we estimate $\bz \approx 74.9 \pm 1.6$\,kG.  This value is a significant fraction of \bs. For a globally dipolar field structure, the ratio of the maximum absolute value of \bz\ to \bs\ does not exceed about 0.3 \citep{Hensetal77,Land88}, so we expect that the maximum \bz\ value observed from a line of sight parallel to the magnetic field axis would be roughly 100\,kG. The actual vale of \bz\ measured suggests that at the time of observation we were observing \wda\ along a line of sight inclined by roughly 40$^\circ$ from the magnetic axis. 

With a single observation, we have no information yet about possible variability in the measured \bz\ and \bs\ magnetic field strengths. If detected, such variations should enable us to determine the WD rotation period, and to obtain a simple model of the surface magnetic field, as was done for WD2047+372 and WD2359--434 \citep{Landetal17}.

\section{The frequency of occurrence of magnetic fields in DA WDs}

\begin{table*}
\caption{Known DA MWDs within 20\,pc volume around Sun}
\label{Tab_da_mwds}
\centering
    \tabcolsep=0.14cm
\begin{tabular}{l l l c r r r r r r c l}
\hline\hline
WD name   & Other name & Class & Distance & \te   & $\log g$ & Mass      & Radius  & Age   & \bs     &\multicolumn{1}{c}{Mag.\ flux}  & References  \\
          &            &       & (pc)     & (K)   & (cgs) & ($M_\odot$)   & ($10^8$\,cm) & (Gyr) & (MG) &
\multicolumn{1}{c}{$\times 10^{-18}$\,MG\,cm$^2$}&       \\
\hline    
 0009+501 &    GJ 1004 &   DAH & 10.87    & 6502 &    8.23  & 0.73       & 7.58    &  3.0  &   0.25  & 0.45  & 1, 2, a \\ 
 0011--721 &   LP 50-73 &   DAH & 18.79    & 6340 &    7.89  & 0.53       & 9.47    &  1.7  &   0.34  & 0.96   & 3, b  \\ 
 0011--134 &    GJ 3016 &   DAH & 18.58    & 5992 &    8.21  & 0.72       & 7.68    &  3.7  &   9.7   & 18.0  & 4, 5, a  \\ 
 0121--429 &  LHS 1243  &   DAH & 18.48    & 6299 &    7.66  & 0.41       & 10.9    &  1.3  &   6.3   &  23.5  & 6, a  \\ 
 0233--242 &  LHS 1421  &   DAH & 18.50    & 5270 &    7.77  & 0.45       & 10.2    &  2.4  &   3.8   & 12.4  & 7, b  \\ 
 0322--019 &    GJ 3223 &  DAZH & 16.91    & 5300 &    8.12  & 0.66       & 8.17    &  5.5  &   0.12  & 0.25  & 8, b  \\ 
 0503--174 &  LHS 1734  &   DAH & 19.35    & 5316 &    7.62  & 0.38       & 11.0    &  1.9  &   4.0   & 15.2  & 4, a  \\ 
 0553+053 &   LHS 212  &   DAH & 7.99     & 5785 &    8.22  & 0.72       & 7.69    &  4.3  &   20    & 37.2  & 9, 10, a  \\ 
 1309+853 &    GJ 3768 &   DAH & 16.47    & 5440 &    8.20  & 0.71       & 7.75    &  5.5  &   4.9   & 9.24  & 11, 12, a  \\ 
 1350--090 & PG 1350-090&   DAH & 19.71    & 9580 &    8.13  & 0.68       & 8.18    &  0.8  &   0.45  & 0.95  & 1, a  \\ 
 1900+705 &    LAWD 73 &   DAP & 12.88   & 11835 &    8.53  & 0.93       & 6.02    &  0.9  &   320   & 364   & 13, 14, 15, 16, a  \\ 
 1953--011 &    LAWD 79 &   DAH & 11.56    & 7868 &    8.23  & 0.73       & 7.66    &  1.6  &   0.50  & 0.92  & 17, 18, 19, a  \\ 
 2047+372 &    GJ 4165 &   DAH & 17.57   & 14600 &    8.33  & 0.82       & 7.11    &  0.4  &   0.06  & 0.095 & 20, 21, b  \\ 
 2105--820 &    LAWD 83 &  DAZP & 16.17    & 9820 &    8.29  & 0.78       & 7.27    &  1.0  &   0.04  & 0.066 & 22, 23, b  \\ 
 2150+591 &            &   DAH & 8.47     & 5095 &    7.98  &            & 9.1     &  5.0  &   0.80  & 2.08  & 24, c  \\ 
 2359--434 &    LAWD 96 &   DAP & 8.34     & 8390 &    8.37  & 0.83       & 6.89    &  1.8  &   0.10  & 0.15  & 25, 21, b  \\ 
\hline
\end{tabular}
\tablefoot{ Sources for magnetic data: (1) \citet{SchmSmit94}, (2) \citet{Valyetal05}, (3) this work, 
(4) \citet{Bergetal92}, (5) \citet{Putn97}, (6) \citet{Subaetal07},
(7) \citet{Vennetal18}, (8) \citet{Farietal11}, (9) \citet{Liebetal75},
(10) \citet{PutnJord95}, (11) \citet{Putn95}, (12) \citet{Putn97},
(13) \citet{Kempetal70}, (14) \citet{LandAnge75}, (15) \citet{Angeetal85},
(16) \citet{BagnLand19}, (17) \citet{Koesetal98}, (18) \citet{Maxtetal00},
(19) \citet{Valyetal08}, (20) \citet{Landetal16}, (21) \citet{Landetal17},
(22) \citet{Koesetal98}, (23) \citet{Landetal12}, (24) \citet{LandBagn19},
(25) \citet{Aznaetal04}. \\ 
Sources for \te, $\log g$, mass, age of WDs: (a) \citet{Giametal12}, (b) \citet{Subaetal17}, (c) \citet{Holletal18} }
\end{table*}

This star is found to be within the 20\,pc volume closest to the Sun. This is an extremely important set of WDs; it is a sample containing almost all the stellar remnants from completed stellar evolution in this volume, and effectively records the results of some 10\,Gyr of completed stellar evolution by all but the most massive stars that have existed in the solar neighbourhood. Because of the importance of this sample, it is of interest to explore in a preliminary way the information contained about the DA stars in the 20\,pc sample in spite of the fact that, at present, the WD sample is still not completely catalogued for spectral class, and the magnetic surveys of the sample are still significantly incomplete \citep{Holletal18}. The statistical properties of magnetic stars in the 20\,pc volume will be examined in greater detail in a forthcoming paper; here we simply sketch some early results.

We take as the fundamental 20\,pc sample of WDs the list provided by \citet{Holletal18}, which is based on careful examination of the recent Gaia Data Release 2 (DR2, Gaia Collaboration \citeyear{Gaiaetal18}). This list (including several WDs without complete Gaia DR2 records) is expected to be at least 95\,\% complete. The list of WDs within the 20\,pc volume presently includes 145 WDs. The new list excludes several stars previously regarded as members of this sample \citep[e.g.][]{Holbetal16}, and now includes a number of WDs not previously thought to be within the sample, as well as (currently) ten stars not previously classified by spectroscopy as WDs (some of which may turn out to be DAs). Consequently, although membership in the latest 20\,pc sample is more firmly established than in earlier versions, the sample is not yet completely definitive, nor fully characterised. 

Recent papers have suggested that the incidence of magnetic fields among WDs may depend strongly on chemical subtype. In particular, it has been suggested that relative to samples including all types of WDs, fields are more frequently found in hot DQ stars \citep{Dufoetal13}, in DAZ stars \citep{KawkVenn14,Kawketal19}, and in DZ stars \citep{Holletal15}. Thus it is important to look at statistics of occurrences of fields in sub-samples of volume-limited samples when this is practical. 

The sub-sample of the 20\,pc WD sample known to have hydrogen-rich outer layers consists of 80 WDs that have been spectroscopically classified be DA, DAZ, DAH or DAP stars. This sub-sample of the full 20\,pc WD sample is particularly interesting. It still contains more than half of the total 20\,pc sample, but because all these stars have evolved into WDs with H-rich outer layers, this sub-sample is probably significantly more homogeneous in origin and evolution than the full 20\,pc sample. Furthermore, all the identified members of the 20\,pc DA sample have reliable upper limits to magnetic fields that might be present, because all have been observed spectroscopically for classification. Such classification spectra usually have sufficient resolution and S/Ns to detect fields with field modulus \bz\ in excess of 1--2\,MG, and in fact fields were discovered by spectroscopy  alone, without spectropolarimetry, in four of the 20\,pc DA stars. 

As will be discussed in more detail in forthcoming publications, most of the stars in the 20\,pc DA sample have been observed spectropolarimetrically. Many of the spectropolarimetric measurements have been published \citep[e.g.][]{Liebetal75,SchmSmit94,SchmSmit95,Putn95,Putn97,Aznaetal04,Jordetal07,Landetal12,Landetal15,Landetal16,BagnLand18,LandBagn19}, but a substantial  number of our measurements are still being prepared for publication. Only about a dozen 20\,pc DA WDs remain unobserved; these are generally faint DA stars with \te\ around 5000\,K and very weak \ha, or in close binary systems.  The spectropolarimetric observations generally have sensitivity enough to detect longitudinal fields \bz\ of a few kG or at worst tens of kG.  This sample is thus not completely observed at the presently attainable level of precision, but is sufficiently thoroughly observed to deserve a preliminary assessment.

The sample of known DA stars within 20\,pc includes 16 well-established DA magnetic stars, with fields (as characterised by measured or estimated values of \bs\ rather than by inferred dipole field strength) ranging from about 40\,kG to over 300\,MG. These 16 stars are listed in Table~\ref{Tab_da_mwds}.  The spectral classes listed in this Table have been given the final letter H if the field is clearly detected in intensity spectra of suitable resolving power, or P if the field is detected only through polarimetry. Distances are taken from the Gaia DR2 (Gaia Collaboration \citeyear{Gaiaetal18}). Values of \te, $\log g$, and mass are adopted from \citet{Giametal12}, \citet{Subaetal17}, or \citet{Holletal18}, as indicated in the footnotes to the Table. The radius of each MWD is computed from the luminosities and \te\ values tabulated by the two main sources of other astrophysical data, except for WD\,2150+591, for which R is deduced from the values of mass and $\log g$ in \citet{Holletal18}. 

 With 16 MWDs out of 80 WDs, the magnetic fraction of DAs is about $(20.0 \pm 5.0)$\,\%. This is well above the overall frequencies generally reported \citep{Liebetal03,Kawketal07,Holbetal16}, but is similar to the frequency suggested by \citet{Jordetal07} for fields in DA stars, and to the global frequency reported (with large uncertainty) for the 13\,pc volume by \citet{KawkVenn14}.

Of the 16 MWDs in the 20\,pc DA sample, nine have $\bs < 1$\,MG, six lie in the range of $1 < \bs < 100$\,MG, and one has a still larger field. It is thus clear that the large fraction of MWDs found in this sample is a result of the large number of weak fields found with sensitive searches in recent years. A sufficiently large number of weak-field MWDs occur to {\em substantially} increase the statistics of occurrence compared to what would be found by a survey sensitive only to MG fields.

Within the full DA sample, half have atmospheres with $\te \leq 7000$\,K and half are hotter than this. This median temperature corresponds to a cooling age of about 1.6\,Gyr \citep{Bergetal95}. The oldest DA WDs in the 20\,pc volume are roughly 5 Gyr old. (We note that older WDs with H-rich atmospheres are almost certainly present in the 20\,pc sample; they have simply cooled to the point of no longer showing atmospheric Balmer lines, and are consequently classified DC stars. These stars are currently not included in our DA sample.) Of the DA MWDs, 10 have $\te \leq 7000$\,kG, while six are hotter than this. The median temperature of the DA MWD sample is about 6300\,K.   The sample of DA MWDs is thus mildly, but not very much, cooler, and older than the general DA sample. 

This high fraction of MWDs among the DA WDs provides a simple test of the hypothesis \citep{Angeetal81} that the MWDs are descendants, through magnetic flux conservation, of the magnetic Ap and Bp stars of the main sequence. Because of their relatively short lives, a substantial fraction of the current DA population is surely descended from A and B main sequence stars of more than about $1.5 M_\odot$ \citep{Holletal15}. These stars currently show a roughly constant magnetic field detection fraction throughout the middle and upper main sequence of about $7 \pm 1$\,\% \citep[e.g.][]{GrunNein15}. If all DA WDs descended from these middle and upper main sequence stars, we would expect to find about 7\,\% of DA WDs to be magnetic. In fact, some of the present DA WDs are certainly descended from less massive stars, so this would be an upper limit to the frequency of occurrence of MWDs. From this argument, and from the observed DA MWD fraction of about 20\,\%, it is clear that even if magnetic main sequence stars do generally become MWDs, there must be other major channels, or else the frequency of magnetic main sequence stars must have been at least three or four times higher in the past than it is today. 

\begin{figure}
\includegraphics*[width=9.3cm,trim=0cm 0cm 0cm  2cm,clip]{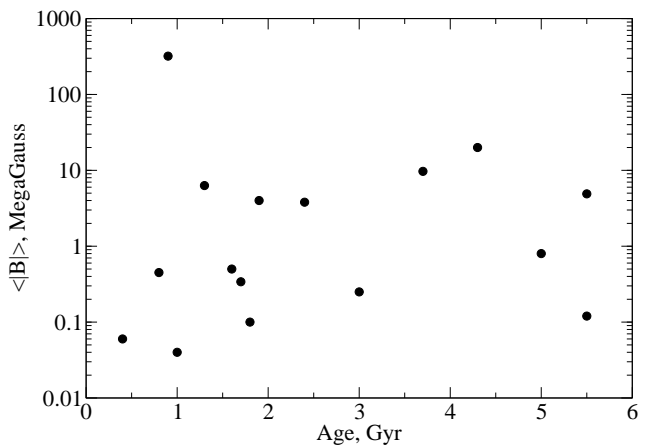}
\caption{\label{Fig_bs_vs_age} 
Mean field modulus of individual DA MWDs within 20\,pc volume-limited sample, as function of WD cooling age.  }
\end{figure}

\begin{figure}
\includegraphics*[width=9.3cm,trim=0cm 0cm 0cm  2cm,clip]{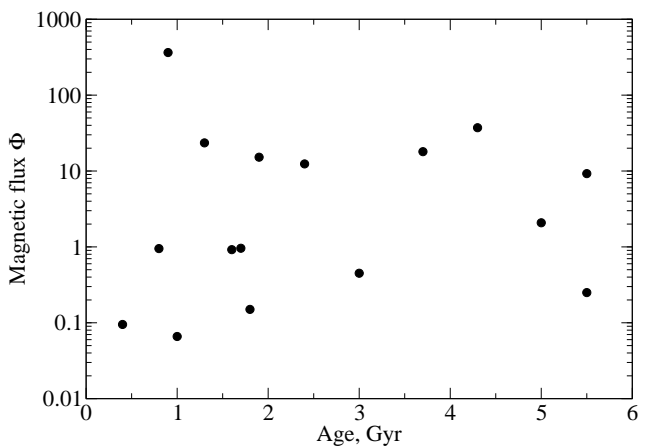}
\caption{\label{Fig_phi_vs_age} 
Estimated magnetic flux of individual DA MWDs within 20\,pc volume-limited sample, as function of cooling age. The magnetic flux is in units of $10^{-18}$\,MG\,cm$^2$, and the age is in Gyr.  }
\end{figure}

The volume-limited 20\,pc DA sample is large enough to be examined for evolutionary trends.  In particular, we may look at this sample to see if clear evidence exists of field decay, which would show up as systematically lower field strengths and magnetic fluxes in older MWDs compared to younger stars.  This kind of study, looking at the statistical variations of field strength and magnetic flux as a function of stellar main sequence age, has clearly revealed that both field strength \bs\ and total magnetic flux in upper main sequence Bp stars decline with time spent on the main sequence \citep{Landetal08,Sikoetal19}. 

 In order to study how magnetic field strength and flux depend on cooling age, we need reasonably accurate cooling times. The cooling ages reported in Table~\ref{Tab_da_mwds} are taken from the same sources as the basic stellar parameters. These cooling ages have been computed for non-magnetic WD models. However, for most of the WDs in our sample, the field is able to suppress or at least greatly reduce the heat flux carried by convection. The inhibition of convection certainly alters the structure of the outer layers of the WDs, and may well affect the spectral diagnostics used to determine basic stellar parameters in ways that have not yet been adequately explored. 

It has been suggested that such major structural change could have very important effects on computed cooling times \citep{Valyetal14}, making the values computed for non-magnetic WDs seriously incorrect for our magnetic DA sample. However, this issue has been explored n depth by \citet{Tremetal15}, who have shown convincingly that convective suppression of convection has no effect whatever on cooling rates for values of \te\ larger than about 6000\,K. Furthermore, the changes to the cooling times for lower values of \te\ \citep[see Figure~6 in][] {Tremetal15}, particularly for values the WD mass around $0.6 M_\odot$ which apply to the coolest MWDs in our sample, are completely inconsequential for \te\ values above 5000\,K. Thus we may safely use the cooling times as computed by \citet{Giametal12} and \citet{Subaetal17}.  

Plots showing \bs\ and the magnetic fluxes in this sample (estimated as $\Phi \sim \pi R^2 \bs$ and tabulated as 'Flux' in Table~\ref{Tab_da_mwds} in units of $10^{-18}$\,MG-cm$^2$) as functions of the cooling age of each MWD are shown in Fig.~\ref{Fig_bs_vs_age} and \ref{Fig_phi_vs_age}. There are no clear trends with age; there is no obvious evidence in this sample for either Ohmic field decay with age, nor for new field generation during cooling. This result is in remarkable contrast to the situation of upper main sequence magnetic stars, in which the flux declines considerably on a time scale of the order of 0.1\,Gy, while for MWDs there is no evidence of decay on a time scale 50 times longer. 
Another interesting feature of these two figures is that it appears that the creation rate of new MWDs per Gyr has not increased remarkably over the 6\,Gyr covered by the figure. 

One may notice that the only MWD with a very large field ($\bs > 10^2$\,MG) occurs at a relatively young age, while there are no corresponding very high field old objects. This is a small-numbers effect, as there are a number of very high-field objects with low values of \te\ and corresponding advanced cooling ages. For example, both G240-72 and G227-35 have fields above 100\,MG, \te\ below 6500\,K, ages of at least about 2\,Gyr, and are within the 20\,pc volume; they are not included in the DA sample because they are not thought to have H-rich atmospheres.

\section{Conclusions}

 In the course of our ongoing spectropolarimetric survey of mostly faint, cool but nearby WDs for weak magnetic fields, we have discovered a magnetic field in \wda. This star is spectral type DA, meaning that the outer layers are H-rich, and the visible spectrum shows only spectral line from the hydrogen Balmer series. Because of the relatively low effective temperature of this WD, about 6340\,K, these lines are rather sharp. It is found from our single FORS polarised spectrum, taken with a resolving power of a little more than 2000, that \ha\ shows clear Zeeman splitting, indicating a mean surface field of about 343\,kG at the time of observation. The \ha\ line also shows strong circular polarisation in the two $\sigma$ components, revealing a longitudinal field \bz\ of about 75\,kG. This field is just slightly too small to have been noticed in classification spectra. 
 
 This star is a member of the very important volume-limited sample of WD stars within 20\,pc of the Sun. As noted above, this sample is effectively a time capsule recording the results of about 10\,Gyr of stellar evolution in the solar neighbourhood. It is by far the most intensely studied volume-limited sample of WDs in searches for magnetic WDs, and contains a significant fraction of all known MWDs with sub-MG magnetic fields. 

The 20\,pc volume contains almost 150 WDs, of which more than half (about 80) are known to have H-rich outer layers. We suspect that the DAs in this volume have probably followed a more homogeneous evolution path that the full sample of 20\,pc WDs. In addition, because all the DAs show at least weak Balmer line spectra, we can detect MG fields in them by conventional low-resolution classification spectroscopy, so that all the classified DAs in the 20\,pc volume either are known to have MG fields, or to have field upper limits of the order of 1--2\,MG. For most of the DAs, spectropolarimetric measurements are also available, with either detected longitudinal fields \bz\ weaker than 1\,MG, or set upper limits of some kG to such fields. Consequently, although the 20\,pc survey is still incomplete, it is worthwhile to carry out a preliminary examination of the statistics of field occurrence among the DA WDs of the 20\,pc sample. 

It is found that 16 WDs, representing about 20\,\% of the DA WDs in the 20\,pc volume, possess detectable magnetic fields of a few kG or more. More than half of these stars have fields below 1\,MG, and were only found to be magnetic as a result of spectropolarimetric observations. It is the addition of these weak-field MWDs to the sample that raises the occurrence frequency of magnetic fields in DA WDs from around 10\% to around 20\,\%. The MWDs in the DA sub-sample of the 20\,pc sample are found to be slightly cooler and older than the typical temperatures and ages of the full DA sample. 
Unlike the magnetic Bp stars of the upper main sequence, in which magnetic flux declines markedly during the $\sim 10^8$\,yr of main sequence evolution, these DA MWDs shown no clear evidence either of field or flux decay during cooling, nor of new field generation, even over a time interval 50 times longer.

\begin{acknowledgements}
We thank the referee, S. O. Kepler, for several helpful suggestions. 
Based on observations made with ESO Telescopes at the
La Silla Paranal Observatory, under programme ID 0102.D-0045.
JDL acknowledges the financial support of the Natural Sciences and
Engineering Research Council of Canada (NSERC), funding reference
number 6377-2016.
\end{acknowledgements}


\bibliography{jdl-sb}

\begin{thebibliography}{57}
\expandafter\ifx\csname natexlab\endcsname\relax\def\natexlab#1{#1}\fi

\bibitem[{{Angel} {et~al.}(1981){Angel}, {Borra}, \& {Landstreet}}]{Angeetal81}
{Angel}, J.~R.~P., {Borra}, E.~F., \& {Landstreet}, J.~D. 1981, \apjs, 45, 457

\bibitem[{{Angel} {et~al.}(1985){Angel}, {Liebert}, \& {Stockman}}]{Angeetal85}
{Angel}, J.~R.~P., {Liebert}, J., \& {Stockman}, H.~S. 1985, \apj, 292, 260

\bibitem[{{Appenzeller} {et~al.}(1998){Appenzeller}, {Fricke}, {F{\"u}rtig},
  {G{\"a}ssler}, {H{\"a}fner}, {Harke}, {Hess}, {Hummel}, {J{\"u}rgens},
  {Kudritzki}, {Mantel}, {Meisl}, {Muschielok}, {Nicklas}, {Rupprecht},
  {Seifert}, {Stahl}, {Szeifert}, \& {Tarantik}}]{Appetal98}
{Appenzeller}, I., {Fricke}, K., {F{\"u}rtig}, W., {et~al.} 1998, The
  Messenger, 94, 1

\bibitem[{{Aznar Cuadrado} {et~al.}(2004){Aznar Cuadrado}, {Jordan},
  {Napiwotzki}, {Schmid}, {Solanki}, \& {Mathys}}]{Aznaetal04}
{Aznar Cuadrado}, R., {Jordan}, S., {Napiwotzki}, R., {et~al.} 2004, \aap, 423,
  1081

\bibitem[{{Bagnulo} {et~al.}(2009){Bagnulo}, {Landolfi}, {Landstreet}, {Landi
  Degl'Innocenti}, {Fossati}, \& {Sterzik}}]{Bagetal09}
{Bagnulo}, S., {Landolfi}, M., {Landstreet}, J.~D., {et~al.} 2009, \pasp, 121,
  993

\bibitem[{{Bagnulo} \& {Landstreet}(2018)}]{BagnLand18}
{Bagnulo}, S. \& {Landstreet}, J.~D. 2018, \aap, 618, A113

\bibitem[{{Bagnulo} \& {Landstreet}(2019)}]{BagnLand19}
{Bagnulo}, S. \& {Landstreet}, J.~D. 2019, \mnras, 486, 4655

\bibitem[{{Bergeron} {et~al.}(1992){Bergeron}, {Ruiz}, \&
  {Leggett}}]{Bergetal92}
{Bergeron}, P., {Ruiz}, M.-T., \& {Leggett}, S.~K. 1992, \apj, 400, 315

\bibitem[{{Bergeron} {et~al.}(1995){Bergeron}, {Wesemael}, \&
  {Beauchamp}}]{Bergetal95}
{Bergeron}, P., {Wesemael}, F., \& {Beauchamp}, A. 1995, \pasp, 107, 1047

\bibitem[{{Brinkworth} {et~al.}(2013){Brinkworth}, {Burleigh}, {Lawrie},
  {Marsh}, \& {Knigge}}]{Brinetal13}
{Brinkworth}, C.~S., {Burleigh}, M.~R., {Lawrie}, K., {Marsh}, T.~R., \&
  {Knigge}, C. 2013, \apj, 773, 47

\bibitem[{{Donati} {et~al.}(1997){Donati}, {Semel}, {Carter}, {Rees}, \&
  {Collier Cameron}}]{Donaetal97}
{Donati}, J.-F., {Semel}, M., {Carter}, B.~D., {Rees}, D.~E., \& {Collier
  Cameron}, A. 1997, \mnras, 291, 658

\bibitem[{{Dufour} {et~al.}(2017){Dufour}, {Blouin}, {Coutu},
  {Fortin-Archambault}, {Thibeault}, {Bergeron}, \& {Fontaine}}]{Dufoetal17}
{Dufour}, P., {Blouin}, S., {Coutu}, S., {et~al.} 2017, in Astronomical Society
  of the Pacific Conference Series, Vol. 509, 20th European White Dwarf
  Workshop, ed. P.-E. {Tremblay}, B.~{Gaensicke}, \& T.~{Marsh}, 3

\bibitem[{{Dufour} {et~al.}(2013){Dufour}, {Vornanen}, {Bergeron}, \&
  {Fontaine}}]{Dufoetal13}
{Dufour}, P., {Vornanen}, T., {Bergeron}, P., \& {Fontaine}, Berdyugin, A.
  2013, in Astronomical Society of the Pacific Conference Series, Vol. 469,
  18th European White Dwarf Workshop., ed. J.~{Krzesi{\'n}ski},
  G.~{Stachowski}, P.~{Moskalik}, \& K.~{Bajan}, 167

\bibitem[{{Farihi} {et~al.}(2011){Farihi}, {Dufour}, {Napiwotzki}, \&
  {Koester}}]{Farietal11}
{Farihi}, J., {Dufour}, P., {Napiwotzki}, R., \& {Koester}, D. 2011, \mnras,
  413, 2559

\bibitem[{{Ferrario} {et~al.}(2015){Ferrario}, {de Martino}, \&
  {G{\"a}nsicke}}]{Ferretal15}
{Ferrario}, L., {de Martino}, D., \& {G{\"a}nsicke}, B.~T. 2015, \ssr, 191, 111

\bibitem[{{Gaia Collaboration} {et~al.}(2018){Gaia Collaboration}, {Brown},
  {Vallenari}, {Prusti}, {de Bruijne}, {Babusiaux}, {Bailer-Jones}, {Biermann},
  {Evans}, {Eyer}, \& et~al.}]{Gaiaetal18}
{Gaia Collaboration}, {Brown}, A.~G.~A., {Vallenari}, A., {et~al.} 2018, \aap,
  616, A1

\bibitem[{{Gary} {et~al.}(2013){Gary}, {Tan}, {Curtis}, {Tristram}, \&
  {Fukui}}]{Garyetal13}
{Gary}, B.~L., {Tan}, T.~G., {Curtis}, I., {Tristram}, P.~J., \& {Fukui}, A.
  2013, Society for Astronomical Sciences Annual Symposium, 32, 71

\bibitem[{{Giammichele} {et~al.}(2012){Giammichele}, {Bergeron}, \&
  {Dufour}}]{Giametal12}
{Giammichele}, N., {Bergeron}, P., \& {Dufour}, P. 2012, \apjs, 199, 29

\bibitem[{{Grunhut} \& {Neiner}(2015)}]{GrunNein15}
{Grunhut}, J.~H. \& {Neiner}, C. 2015, in IAU Symposium, Vol. 305, Polarimetry,
  ed. K.~N. {Nagendra}, S.~{Bagnulo}, R.~{Centeno}, \& M.~{Jes{\'u}s
  Mart{\'{\i}}nez Gonz{\'a}lez}, 53--60

\bibitem[{{Hensberge} {et~al.}(1977){Hensberge}, {van Rensbergen}, {Goossens},
  \& {Deridder}}]{Hensetal77}
{Hensberge}, H., {van Rensbergen}, W., {Goossens}, M., \& {Deridder}, G. 1977,
  \aap, 61, 235

\bibitem[{{Holberg} {et~al.}(2016){Holberg}, {Oswalt}, {Sion}, \&
  {McCook}}]{Holbetal16}
{Holberg}, J.~B., {Oswalt}, T.~D., {Sion}, E.~M., \& {McCook}, G.~P. 2016,
  \mnras, 462, 2295

\bibitem[{{Hollands} {et~al.}(2015){Hollands}, {G{\"a}nsicke}, \&
  {Koester}}]{Holletal15}
{Hollands}, M.~A., {G{\"a}nsicke}, B.~T., \& {Koester}, D. 2015, \mnras, 450,
  681

\bibitem[{{Hollands} {et~al.}(2018){Hollands}, {Tremblay}, {G{\"a}nsicke},
  {Gentile-Fusillo}, \& {Toonen}}]{Holletal18}
{Hollands}, M.~A., {Tremblay}, P.-E., {G{\"a}nsicke}, B.~T., {Gentile-Fusillo},
  N.~P., \& {Toonen}, S. 2018, \mnras, 480, 3942

\bibitem[{{Jordan} {et~al.}(2007){Jordan}, {Aznar Cuadrado}, {Napiwotzki},
  {Schmid}, \& {Solanki}}]{Jordetal07}
{Jordan}, S., {Aznar Cuadrado}, R., {Napiwotzki}, R., {Schmid}, H.~M., \&
  {Solanki}, S.~K. 2007, \aap, 462, 1097

\bibitem[{{Kawka} \& {Vennes}(2014)}]{KawkVenn14}
{Kawka}, A. \& {Vennes}, S. 2014, \mnras, 439, L90

\bibitem[{{Kawka} {et~al.}(2019){Kawka}, {Vennes}, {Ferrario}, \&
  {Paunzen}}]{Kawketal19}
{Kawka}, A., {Vennes}, S., {Ferrario}, L., \& {Paunzen}, E. 2019, \mnras, 482,
  5201

\bibitem[{{Kawka} {et~al.}(2007){Kawka}, {Vennes}, {Schmidt}, {Wickramasinghe},
  \& {Koch}}]{Kawketal07}
{Kawka}, A., {Vennes}, S., {Schmidt}, G.~D., {Wickramasinghe}, D.~T., \&
  {Koch}, R. 2007, \apj, 654, 499

\bibitem[{{Kemp} {et~al.}(1970){Kemp}, {Swedlund}, {Landstreet}, \&
  {Angel}}]{Kempetal70}
{Kemp}, J.~C., {Swedlund}, J.~B., {Landstreet}, J.~D., \& {Angel}, J.~R.~P.
  1970, \apjl, 161, L77

\bibitem[{{Koester} {et~al.}(1998){Koester}, {Dreizler}, {Weidemann}, \&
  {Allard}}]{Koesetal98}
{Koester}, D., {Dreizler}, S., {Weidemann}, V., \& {Allard}, N.~F. 1998, \aap,
  338, 612

\bibitem[{{Landstreet}(1988)}]{Land88}
{Landstreet}, J.~D. 1988, \apj, 326, 967

\bibitem[{{Landstreet} \& {Angel}(1975)}]{LandAnge75}
{Landstreet}, J.~D. \& {Angel}, J.~R.~P. 1975, \apj, 196, 819

\bibitem[{{Landstreet} \& {Bagnulo}(2019)}]{LandBagn19}
{Landstreet}, J.~D. \& {Bagnulo}, S. 2019, \aap, 623, A46

\bibitem[{{Landstreet} {et~al.}(2016){Landstreet}, {Bagnulo}, {Martin}, \&
  {Valyavin}}]{Landetal16}
{Landstreet}, J.~D., {Bagnulo}, S., {Martin}, A., \& {Valyavin}, G. 2016, \aap,
  591, A80

\bibitem[{{Landstreet} {et~al.}(2017){Landstreet}, {Bagnulo}, {Valyavin}, \&
  {Valeev}}]{Landetal17}
{Landstreet}, J.~D., {Bagnulo}, S., {Valyavin}, G., \& {Valeev}, A.~F. 2017,
  \aap, 607, A92

\bibitem[{{Landstreet} {et~al.}(2012){Landstreet}, {Bagnulo}, {Valyavin},
  {Fossati}, {Jordan}, {Monin}, \& {Wade}}]{Landetal12}
{Landstreet}, J.~D., {Bagnulo}, S., {Valyavin}, G.~G., {et~al.} 2012, \aap,
  545, A30

\bibitem[{{Landstreet} {et~al.}(2015){Landstreet}, {Bagnulo}, {Valyavin},
  {Gadelshin}, {Martin}, {Galazutdinov}, \& {Semenko}}]{Landetal15}
{Landstreet}, J.~D., {Bagnulo}, S., {Valyavin}, G.~G., {et~al.} 2015, \aap,
  580, A120

\bibitem[{{Landstreet} {et~al.}(2008){Landstreet}, {Silaj}, {Andretta},
  {Bagnulo}, {Berdyugina}, {Donati}, {Fossati}, {Petit}, {Silvester}, \&
  {Wade}}]{Landetal08}
{Landstreet}, J.~D., {Silaj}, J., {Andretta}, V., {et~al.} 2008, \aap, 481, 465

\bibitem[{{Liebert} {et~al.}(1975){Liebert}, {Angel}, \&
  {Landstreet}}]{Liebetal75}
{Liebert}, J., {Angel}, J.~R.~P., \& {Landstreet}, J.~D. 1975, \apjl, 202, L139

\bibitem[{{Liebert} {et~al.}(2003){Liebert}, {Bergeron}, \&
  {Holberg}}]{Liebetal03}
{Liebert}, J., {Bergeron}, P., \& {Holberg}, J.~B. 2003, \aj, 125, 348

\bibitem[{{Mathys}(1989)}]{Math89}
{Mathys}, G. 1989, \fcp, 13, 143

\bibitem[{{Maxted} {et~al.}(2000){Maxted}, {Ferrario}, {Marsh}, \&
  {Wickramasinghe}}]{Maxtetal00}
{Maxted}, P.~F.~L., {Ferrario}, L., {Marsh}, T.~R., \& {Wickramasinghe}, D.~T.
  2000, \mnras, 315, L41

\bibitem[{{Preston}(1970)}]{Pres70}
{Preston}, G.~W. 1970, \apjl, 160, L143

\bibitem[{{Putney}(1995)}]{Putn95}
{Putney}, A. 1995, \apjl, 451, L67

\bibitem[{{Putney}(1997)}]{Putn97}
{Putney}, A. 1997, \apjs, 112, 527

\bibitem[{{Putney} \& {Jordan}(1995)}]{PutnJord95}
{Putney}, A. \& {Jordan}, S. 1995, \apj, 449, 863

\bibitem[{{Schmidt} \& {Smith}(1994)}]{SchmSmit94}
{Schmidt}, G.~D. \& {Smith}, P.~S. 1994, \apjl, 423, L63

\bibitem[{{Schmidt} \& {Smith}(1995)}]{SchmSmit95}
{Schmidt}, G.~D. \& {Smith}, P.~S. 1995, \apj, 448, 305

\bibitem[{{Sikora} {et~al.}(2019){Sikora}, {Wade}, {Power}, \&
  {Neiner}}]{Sikoetal19}
{Sikora}, J., {Wade}, G.~A., {Power}, J., \& {Neiner}, C. 2019, \mnras, 483,
  3127

\bibitem[{{Subasavage} {et~al.}(2007){Subasavage}, {Henry}, {Bergeron},
  {Dufour}, {Hambly}, \& {Beaulieu}}]{Subaetal07}
{Subasavage}, J.~P., {Henry}, T.~J., {Bergeron}, P., {et~al.} 2007, \aj, 134,
  252

\bibitem[{{Subasavage} {et~al.}(2017){Subasavage}, {Jao}, {Henry}, {Harris},
  {Dahn}, {Bergeron}, {Dufour}, {Dunlap}, {Barlow}, {Ianna}, {L{\'e}pine}, \&
  {Margheim}}]{Subaetal17}
{Subasavage}, J.~P., {Jao}, W.-C., {Henry}, T.~J., {et~al.} 2017, \aj, 154, 32

\bibitem[{{Toonen} {et~al.}(2017){Toonen}, {Hollands}, {G{\"a}nsicke}, \&
  {Boekholt}}]{Tooetal17}
{Toonen}, S., {Hollands}, M., {G{\"a}nsicke}, B.~T., \& {Boekholt}, T. 2017,
  \aap, 602, A16

\bibitem[{{Tout} {et~al.}(2008){Tout}, {Wickramasinghe}, {Liebert}, {Ferrario},
  \& {Pringle}}]{Toutetal08}
{Tout}, C.~A., {Wickramasinghe}, D.~T., {Liebert}, J., {Ferrario}, L., \&
  {Pringle}, J.~E. 2008, \mnras, 387, 897

\bibitem[{{Tremblay} {et~al.}(2015){Tremblay}, {Fontaine}, {Freytag},
  {Steiner}, {Ludwig}, {Steffen}, {Wedemeyer}, \& {Brassard}}]{Tremetal15}
{Tremblay}, P.-E., {Fontaine}, G., {Freytag}, B., {et~al.} 2015, \apj, 812, 19

\bibitem[{{Valyavin} {et~al.}(2005){Valyavin}, {Bagnulo}, {Monin}, {Fabrika},
  {Lee}, {Galazutdinov}, {Wade}, \& {Burlakova}}]{Valyetal05}
{Valyavin}, G., {Bagnulo}, S., {Monin}, D., {et~al.} 2005, \aap, 439, 1099

\bibitem[{{Valyavin} {et~al.}(2014){Valyavin}, {Shulyak}, {Wade}, {Antonyuk},
  {Zharikov}, {Galazutdinov}, {Plachinda}, {Bagnulo}, {Fox Machado}, \&
  {Alvarez}}]{Valyetal14}
{Valyavin}, G., {Shulyak}, D., {Wade}, G.~A., {et~al.} 2014, \nat, 515, 88

\bibitem[{{Valyavin} {et~al.}(2008){Valyavin}, {Wade}, {Bagnulo}, {Szeifert},
  {Landstreet}, {Han}, \& {Burenkov}}]{Valyetal08}
{Valyavin}, G., {Wade}, G.~A., {Bagnulo}, S., {et~al.} 2008, \apj, 683, 466

\bibitem[{{Vennes} {et~al.}(2018){Vennes}, {Kawka}, {Ferrario}, \&
  {Paunzen}}]{Vennetal18}
{Vennes}, S., {Kawka}, A., {Ferrario}, L., \& {Paunzen}, E. 2018, Contributions
  of the Astronomical Observatory Skalnate Pleso, 48, 307

\end{thebibliography}

\end{document}